\documentclass{ws-p8-50x6-00}
\usepackage{
feynmf}
\newcommand{\tx}{\tilde{x}}
\newcommand{\tp}{\tilde{p}}
\newcommand{\mean}[1]{\left\langle#1\right\rangle_0^{p_0x_0}}
\newcommand{\meanfree}[1]{\left\langle#1\right\rangle_{\rm free}^{x_0}}
\newcommand{\cum}[1]{\left\langle#1\right\rangle_{0,c}^{p_0x_0}}
\newcommand{\cumfree}[1]{\left\langle#1\right\rangle_{{\rm free},c}^{x_0}}
\newcommand{\otau}{\overline{\tau}}
\def\setval{\fmfset{wiggly_len}{2mm}\fmfset{arrow_len}{2mm}\fmfset{dash_len}{1.5mm}
\fmfpen{0.125mm}\fmfset{dot_size}{1thick}}
\newcommand{\vertex}[1]{\fmfv{decor.shape=pentagram, decor.filled=1, decor.size=2thick}{#1}}
\begin{document}
\title{Effective Classical Hamiltonian from Perturbatively Defined Path Integral}
\author{M. Bachmann}
\address{Freie Universit\"at Berlin, Institut f\"ur Theoretische Physik,\\
Arnimallee 14, 14195 Berlin, Germany\\
E-mail: mbach@physik.fu-berlin.de}
\maketitle
\abstracts{
Introducing a perturbative definition, phase 
space path integrals can be calculated without slicing.
This leads to a short-time expansion of the quantum-mechanical
path amplitude, or a high-temperature expansion of the unnormalized
density matrix, respectively. We use the proposed formalism to calculate
the effective classical Hamiltonian for the harmonic oscillator. 
}
\section{Introduction}
Path integrals are usually evaluated by time-slicing\cite{PI}, since the
continuum definition is mathematically problematic. This becomes obvious for
physical systems with nontrivial metric, where reparametrization invariance
seems to be violated.

An alternative method of calculation was proposed on the basis
of a perturbative definition of path integrals in 
{\it configuration space}\cite{cherv} motivated by perturbative procedures in quantum
field theory. In this approach, the path integral 
of any system is expanded about the exactly known solution for the free particle
in powers of the coupling constant of the potential.

The extension of this definition to functional integrals in {\it phase space}
is presented in this paper by treating the complete Hamilton function as perturbation,
which leads to a short-time expansion of the quantum-mechanical
path amplitude. In Euclidean space, the density matrix is obtained as a high-temperature 
expansion. By simple resummation, this series can
be turned into an expansion in powers of the coupling constant 
of the potential described above.  
As will be shown in the sequel, the knowledge of an exactly known nontrivial path 
integral as that of a free particle is, however, {\it not} required. Thus, the 
perturbative definition presented here is more general. It reproduces the expansion around
the free particle by a simple re-summation.

The method is then applied to calculate the effective classical Hamiltonian of the
harmonic oscillator $H_{\omega,{\rm eff}}(p_0,x_0)$ by exactly summing up the perturbation series. 
This quantity is related to the quantum statistical partition 
function via the classical looking phase space integral
\begin{equation}
\label{partfunc}
Z_\omega=\int\frac{dx_0dp_0}{2\pi\hbar}\,\exp\left\{-\beta H_{\omega,{\rm eff}}(p_0,x_0)\right\},
\end{equation}
where $\beta=1/k_BT$ is the inverse thermal energy.
\section{Perturbative Definition of the Path Integral for the Density Matrix}
Slicing the interval $[0,\hbar\beta]$ into $N+1$ pieces of width $\varepsilon=\hbar\beta/(N+1)$,
the unnormalized density matrix can be expressed in the continuum limit by\cite{PI}
\begin{eqnarray}
  \label{densslic}
  \tilde{\varrho}(x_b,x_a)&=&\lim_{N\to\infty}\prod\limits_{n=1}^N
\left[\int_{-\infty}^\infty dx_n\right]
\prod\limits_{n=1}^{N+1}\left[\int_{-\infty}^\infty \frac{dp_n}{2\pi\hbar}
\,e^{ip_n(x_n-x_{n-1})}\right]\nonumber\\
&&\times\exp\left\{-\varepsilon 
\sum\limits_{n=1}^{N+1}\,H(p_n,x_n)/\hbar\right\},
\end{eqnarray}
where $x_a=x_0$ and $x_b=x_{N+1}$ are the fixed end points of the path.
Expanding the last exponential in powers of $\varepsilon/\hbar$, we recognize that the
zeroth order contribution to the density matrix (\ref{densslic}) is an 
{\it infinite product of $\delta$-functions} due to the identity
\begin{equation} 
\label{deltafunc}
\int_{-\infty}^\infty \frac{dp_n}{2\pi\hbar}\,e^{ip_n(x_n-x_{n-1})}=\delta(x_n-x_{n-1}).
\end{equation}
This infinite product simply reduces to
\begin{equation}
\label{infprod}
\lim_{N\to \infty}\int\limits_{-\infty}^\infty dx_N\cdots dx_1\,
\delta(x_{N+1}-x_N)\cdots\delta(x_2-x_1)\delta(x_1-x_0)
=\delta(x_b-x_a).
\end{equation}
Note that this is the unperturbed contribution to the unnormalized
density matrix (\ref{densslic}) which 
was obtained without solving a nontrivial path integral. Thus, the phase space path integral 
for the unnormalized density matrix (\ref{densslic}) can be perturbatively defined as
\begin{eqnarray}
  \label{denspert}
  \tilde{\varrho}(x_b,x_a)&=&\delta(x_b-x_a)
+\sum\limits_{n=1}^\infty\frac{(-1)^n}{\hbar^n n!}
\int_0^{\hbar\beta}d\tau_1\cdots\int_0^{\hbar\beta}d\tau_n\nonumber\\
&&\times\left\langle
H(p(\tau_1),x(\tau_1))\cdots H(p(\tau_n),x(\tau_n)) \right\rangle_0
\end{eqnarray}
with expectation values
\begin{equation}
\langle\cdots\rangle_0=\lim_{N\to\infty}\prod\limits_{n=1}^N
\left[\int_{-\infty}^\infty dx_n\right]
\prod\limits_{n=1}^{N+1}\left[\int_{-\infty}^\infty \frac{dp_n}{2\pi\hbar}
\,\cdots e^{ip_n(x_n-x_{n-1})}\right].
\end{equation}
These expectation values are usually reexpressed by Feynman diagrams. This is
possible for polynomial as well as nonpolynomial functions of momentum and position\cite{corr}. 
\section{Restricted Partition Function and Two-Point Correlations}
The trace over the unnormalized density matrix (\ref{denspert}) of our unperturbed system with 
vanishing Hamiltonian $H(p,x)=0$ leads to a diverging partition function. It is also
obvious that the classical partition function diverges with the phase space volume.
The regularization of these divergences is possible by excluding  
from the phase space path integral
the zero frequency
fluctuations $x_0$ and $p_0$ of the Fourier decomposition of the periodic path $x(\tau)$ and 
momentum $p(\tau)$, respectively\cite{PI,mag}. 
Therefore we express the quantum statistical partition function of any system by 
\begin{equation}
  \label{clpart}
  Z=\int\frac{dx_0dp_0}{2\pi\hbar}\,Z^{p_0x_0},
\end{equation}
where the restricted partition function was introduced as the Boltzmann factor 
of the effective classical Hamiltonian:
\begin{eqnarray}
\label{eq:part}
&&\hspace{-5mm}Z^{p_0x_0}\equiv \exp\left\{-\beta H_{\rm eff}(p_0,x_0)\right\}=
2\pi\hbar\oint {\cal D}x{\cal D}p\,\delta(x_0-\overline{x})
\delta(p_0-\overline{p})\nonumber\\
&&\hspace{-2mm}\times\exp\left\{-\frac{1}{\hbar}\int_0^{\hbar\beta}d\tau\,
\left[-i(p(\tau)-p_0)\frac{\partial}{\partial \tau}(x(\tau)-x_0)+
H(p(\tau),x(\tau))\right]  \right\},
\end{eqnarray}
with the measure 
\begin{equation}
\oint {\cal D}x{\cal D}p=\lim_{N\to\infty} \prod\limits_{n=1}^{N+1}\left[\int_{-\infty}^\infty 
\frac{dx_ndp_n}{2\pi\hbar} \right]
\end{equation}
and temporal mean values $\overline{x}=\int_0^{\hbar\beta}d\tau x(\tau)/\hbar\beta$ and
$\overline{p}=\int_0^{\hbar\beta}d\tau p(\tau)/\hbar\beta$.
As in the preceding section, the unperturbed system may have a vanishing Hamiltonian, i.e. 
$H=0$. The
calculation of the restricted partition function $Z^{p_0x_0}_0$ for this system, i.e. evaluating
the path integral (\ref{eq:part}) for $H=0$,
is in the same sense {\it trivial} as for its density matrix, since it also reduces 
to a cancellation of $\delta$-functions and yields $Z^{p_0x_0}_0=1$.

In what follows, we concentrate on the correlation functions of position and
momentum dependent quantitites. For this purpose it is convenient to introduce
the generating functional
\begin{eqnarray}
  \label{eq:func}
&&\hspace{-7mm}Z^{p_0x_0}_0[j,v]=2\pi\hbar\oint {\cal D}x{\cal D}p\,\delta(x_0-\overline{x})
\delta(p_0-\overline{p})\nonumber\\
&&\hspace{7mm}\times\exp\left\{ -\frac{1}{\hbar}\int_0^{\hbar\beta}d\tau\,
\left[-i\tp(\tau)\frac{\partial}{\partial \tau}\tx(\tau)+ j(\tau)\tx(\tau)
+v(\tau)\tp(\tau)\right]\right\}
\end{eqnarray}
with abbreviations $\tx(\tau)=x(\tau)-x_0$ and $\tp(\tau)=p(\tau)-p_0$. The calculation
yields
\begin{equation}
  \label{eq:func2}
  Z^{p_0x_0}_0[j,v]=\exp\left\{\frac{1}{\hbar^2}\int_0^{\hbar\beta}d\tau\int_0^{\hbar\beta} 
d\tau'\,j(\tau)G^{p_0x_0}(\tau,\tau')v(\tau')\right\},
\end{equation}
where the periodic Green function is
\begin{eqnarray}
  \label{eq:prop}
  G^{p_0x_0}(\tau,\tau')&=&-\frac{i}{2\beta}\left\{2(\tau-\tau')-
\hbar\beta\left[\Theta(\tau-\tau')-\Theta(\tau'-\tau)\right] \right\}\nonumber\\
&=&\frac{2i}{\beta}\sum\limits_{m=1}^\infty\,\frac{\sin \omega_m(\tau-\tau')}{\omega_m}.
\end{eqnarray}
In the last line we have given the Fourier decomposition with respect to Matsubara frequencies
$\omega_m=2\pi m/\hbar\beta$ omitting the zero mode. Note that 
$G^{p_0x_0}(\tau,\tau')=-G^{p_0x_0}(\tau',\tau)$. These Green functions possess
an interesting scaling property: Substituting $\overline{\tau}\equiv \tau/\beta$, the Green
function becomes {\it independent} of $\beta$:
\begin{equation}
  \label{eq:scalprop}
  G^{p_0x_0}(\otau,\otau')=-\frac{i}{2}\left\{2(\otau-\otau')-
\hbar\left[\Theta(\otau-\otau')-\Theta(\otau'-\otau)\right] \right\}.
\end{equation}
Introducing expectation values as
\begin{equation}
  \label{eq:expect}
  \mean{\cdots}=2\pi\hbar\oint {\cal D}x{\cal D}p\,\delta(x_0-\overline{x})
\delta(p_0-\overline{p})\,\cdots \exp\left\{ \frac{1}{\hbar}\int_0^{\hbar\beta}d\tau\,
i\tp(\tau)\frac{\partial}{\partial \tau}\tx(\tau)\right\},
\end{equation}
the two-point functions are obtained from functional (\ref{eq:func}) by performing 
appropriate functional derivatives with respect to $j(\tau)$ and $v(\tau)$, respectively:
\begin{eqnarray}
\label{eq:corr1}
\mean{\tx(\tau)\tx(\tau')}&=&0,\\
\label{eq:corr2}
\mean{\tx(\tau)\tp(\tau')}&=&G^{p_0x_0}(\tau,\tau'),\\
\label{eq:corr3}
\mean{\tp(\tau)\tp(\tau')}&=&0.
\end{eqnarray}
The result is that only {\it mixed} position-momentum correlations do not vanish.
\section{Perturbative Expansion for Effective Classical Hamiltonian}
Expanding the restricted partition function (\ref{eq:part}) about a vanishing Hamiltonian,
\begin{eqnarray}
  \label{eq:expand}
Z^{p_0x_0}&=&1+\sum\limits_{n=1}^\infty \frac{(-1)^n}{\hbar^n n!}\int_0^{\hbar\beta}d\tau_1\cdots
\int_0^{\hbar\beta}d\tau_n\,\nonumber\\
&&\times\mean{H(p(\tau_1),x(\tau_1))\cdots H(p(\tau_n),x(\tau_n))},
\end{eqnarray}
rewriting this into a cumulant expansion, and utilizing the relation (\ref{eq:part}) 
between restricted partition function and effective classical Hamiltonian, we obtain
\begin{eqnarray}
  \label{eq:ham}
  H_{\rm eff}(p_0,x_0)&=&\frac{1}{\beta}\sum\limits_{n=1}^\infty \frac{(-1)^{n+1}}{\hbar^n n!}
\int_0^{\hbar\beta}d\tau_1\cdots \int_0^{\hbar\beta}d\tau_n\,\nonumber\\
&&\times\cum{H(p(\tau_1),x(\tau_1))\cdots H(p(\tau_n),x(\tau_n))}.
\end{eqnarray}
Using Wick's rule, all correlation functions can be expressed in terms of products
of two-point functions. Since only mixed two-point functions (\ref{eq:prop}) can lead to 
nonvanishing contributions to the effective classical Hamiltonian, we use the rescaled
version (\ref{eq:scalprop}) of the Green function. The scaling transformation gives
a factor $\beta$ from each of the $n$ integral measures. Thus the expansion (\ref{eq:ham})
is a {\it high-temperature} expansion of the effective classical Hamiltonian:
\begin{eqnarray}
  \label{eq:scalham}
  H_{\rm eff}(p_0,x_0)&=&\sum\limits_{n=1}^\infty \beta^{n-1}\frac{(-1)^{n+1}}{\hbar^n n!}
\int_0^{\hbar}d\otau_1\cdots \int_0^{\hbar}d\otau_n\,\nonumber\\
&&\times\cum{H(p(\otau_1),x(\otau_1))\cdots H(p(\otau_n),x(\otau_n))}.
\end{eqnarray}  
For the following considerations it is useful to assume the Hamilton function to be 
of standard form: $H(p(\otau),x(\otau))=p^2(\otau)/2M+gV(x(\otau))$. Here we have introduced
the coupling constant $g$ of the potential. Defining the functionals 
$a[p]=\int_0^{\hbar}d\otau\,p^2(\otau)/2M$ and $b[x]=\int_0^{\hbar}d\otau\,V(x(\otau))$,
the high-temperature expansion (\ref{eq:scalham}) is expressed as
\begin{equation}
  \label{eq:binham}
  H_{\rm eff}(p_0,x_0)=\sum\limits_{n=1}^\infty\beta^{n-1}\frac{(-1)^{n+1}}{n! \hbar^n}
\sum\limits_{k=0}^{n}g^k\left(\begin{array}{c}n \\ k \end{array}\right)\cum{a^{n-k}[p]b^k[x]}.
\end{equation}
In the sequel we point out how this high-temperature expansion is connected
with an expansion in powers of the coupling constant $g$ of the potential. 
\section{High-Temperature Versus Weak-Coupling Expansion}
In the preceding section it was shown that the perturbative expansion about a vanishing
Hamiltonian leads to a perturbative series in powers of the inverse temperature in a
natural manner. Now we elaborate the relation to a perturbative expansion in powers
of the coupling constant $g$ of the potential. Changing the order of summation in 
Eq.~(\ref{eq:binham}), one obtains
\begin{equation}
  H_{\rm eff}(p_0,x_0)=\sum\limits_{k=0}^\infty g^k\sum\limits_{n=0}^\infty \beta^{n+k-1}
\left(\begin{array}{c}n+k \\ k \end{array}\right)\frac{(-1)^{n+k+1}}{(n+k)!\hbar^{n+k}}
\cum{a^n[p]b^k[x]}+\frac{1}{\beta},
\end{equation}
which is rewritten after explicitly evaluating the ($n=0$)- and ($k=0$)-contributions:
\begin{eqnarray}
  \label{eq:free1}
H_{\rm eff}(p_0,x_0)&=&\frac{p_0^2}{2M}+gV(x_0)+\frac{1}{\beta}
\sum\limits_{k=1}^\infty g^k\sum\limits_{n=1}^\infty \frac{(-1)^{n+k+1}}{n!k!\hbar^{n+k}(2M)^n}
\nonumber\\
&&\times\int_0^{\hbar\beta}d\tau_1\cdots\int_0^{\hbar\beta}d\tau_k
\int_0^{\hbar\beta}d\tau_{k+1}\cdots
\int_0^{\hbar\beta}d\tau_{k+n}\nonumber\\
&&\times\cum{V(x(\tau_1))\cdots V(x(\tau_k))p^2(\tau_{k+1})\cdots
p^2(\tau_{k+n})}
\end{eqnarray}
In this expression, we have inversed the scaling transformation and used that
$\int_0^{\hbar\beta}d\tau\,\cum{p^2(\tau)}= \hbar\beta p_0^2$ and 
$\int_0^{\hbar\beta}d\tau\,\cum{V(x(\tau))}= \hbar\beta V(x_0)$. Higher-order
expectations of functions only depending on $x$ {\it or} $p$ are zero. This is due to
the vanishing of expectations of functions of $\tx$ or $\tp$ which are decomposed 
into products of two-point functions (\ref{eq:corr1}) and (\ref{eq:corr3}). All other possible 
contributions are non-connected.

Note that the expansion (\ref{eq:free1}) is equal to the expansion about the free particle
\begin{eqnarray}
H_{\rm eff}(p_0,x_0)&=&\frac{p_0^2}{2M}+gV(x_0)+\frac{1}{\beta}
\sum\limits_{k=1}^\infty g^k\frac{(-1)^{k+1}}{k!\hbar^{k}}
\int_0^{\hbar\beta}d\tau_1\cdots\int_0^{\hbar\beta}d\tau_k\nonumber\\
&&\times \cumfree{V(x(\tau_1))\cdots V(x(\tau_k))},
\label{eq:free2}  
\end{eqnarray}
where the new cumulants are formed from expectation values 
\begin{eqnarray}
  \label{eq:free3}
&&\hspace{-5mm}\meanfree{\cdots}=2\pi\hbar\oint {\cal D}x{\cal D}p\delta(x_0-\overline{x})
\delta(p_0-\overline{p})\cdots \nonumber\\
&&\times\exp\left\{-\frac{1}{\hbar}
\int_0^{\hbar\beta}d\tau\,\left[-i(p(\tau)-p_0)\frac{\partial}{\partial \tau}(x(\tau)-x_0) +
\frac{1}{2M}(p(\tau)-p_0)^2\right] \right\}.\nonumber\\
\quad
\end{eqnarray}
\section{Effective Classical Hamiltonian of Harmonic Oscillator}
In this section, we calculate the effective classical Hamiltonian for the harmonic
oscillator 
\begin{equation}
H_\omega(p,x)=\frac{p^2}{2M}+\frac{1}{2}M\omega^2x^2
\end{equation}
by an exact resummation of the high-temperature expansion (\ref{eq:binham}). 
For systematically expressing the terms of this expansion,
it is useful to introduce the following Feynman rules:
\begin{fmffile}{mbprop}
\begin{eqnarray}
\label{rule0a}
\parbox{30mm}{\centerline{
\begin{fmfgraph*}(20,8)
\setval
\fmfleft{v1}
\fmfright{v2}
\fmf{wiggly}{v2,v1}
\fmflabel{$\otau_1$}{v1}
\fmflabel{$\otau_2$}{v2}
\end{fmfgraph*}
}} &\equiv& \mean{p(\otau_1)p(\otau_2)}=p_0^2,\\
\label{rule0b}
\parbox{30mm}{\centerline{
\begin{fmfgraph*}(20,8)
\setval
\fmfleft{v1}
\fmfright{v2}
\fmf{plain}{v2,v1}
\fmflabel{$\otau_1$}{v1}
\fmflabel{$\otau_2$}{v2}
\end{fmfgraph*}
}} &\equiv& \mean{x(\otau_1)x(\otau_2)}=x_0^2,\\
\label{rule1a}
\parbox{30mm}{\centerline{
\begin{fmfgraph*}(20,8)
\setval
\fmfleft{v1}
\fmfright{v2}
\fmf{dashes_arrow}{v2,v1}
\fmflabel{$\otau_1$}{v1}
\fmflabel{$\otau_2$}{v2}
\end{fmfgraph*}
}} &\equiv& \mean{x(\otau_1)p(\otau_2)}=G^{p_0x_0}(\otau_1,\otau_2)+x_0p_0,\\
\label{rule1b}
\parbox{30mm}{\centerline{
\begin{fmfgraph*}(20,8)
\setval
\fmfleft{v1}
\fmfright{v2}
\fmf{dashes_arrow}{v1,v2}
\fmflabel{$\otau_1$}{v1}
\fmflabel{$\otau_2$}{v2}
\end{fmfgraph*}
}} &\equiv& \mean{p(\otau_1)x(\otau_2)}=-G^{p_0x_0}(\otau_1,\otau_2)+p_0x_0,\\
\label{rule2}
\parbox{30mm}{\centerline{
\begin{fmfgraph*}(15,8)
\setval
\fmfleft{v1}
\fmfright{v2}
\fmf{wiggly}{v2,v1}
\fmflabel{$\otau$}{v1}
\vertex{v2}
\end{fmfgraph*}
}} &\equiv& \mean{p(\otau)}=p_0,\\
\label{rule3}
\parbox{30mm}{\centerline{
\begin{fmfgraph*}(15,8)
\setval
\fmfleft{v1}
\fmfright{v2}
\fmf{plain}{v2,v1}
\fmflabel{$\otau$}{v1}
\vertex{v2}
\end{fmfgraph*}
}} &\equiv& \mean{x(\otau)}=x_0,\\
\label{rule4}
\parbox{30mm}{\centerline{
\begin{fmfgraph}(15,8)
\setval
\fmfforce{0.5w,0.5h}{v1}
\fmfdot{v1}
\end{fmfgraph}
}} &\equiv& \int_0^{\hbar}d\otau,
\end{eqnarray}
where the current-like expectations in (\ref{rule2}) and (\ref{rule3}) arise from
$\mean{\tp(\otau)}=0$ and $\mean{\tx(\otau)}=0$, respectively. In order to simplify the 
investigation of the expectation values in the high-temperature
expansion of the effective classical Hamiltonian~(\ref{eq:scalham}), we also
define operational subgraphs which are useful to systematically construct Feynman
diagrams:
\begin{eqnarray}
\label{rule5}
\parbox{30pt}{\centerline{
\begin{fmfgraph}(20,8)
\setval
\fmfleft{v1}
\fmfright{v3}
\fmfforce{0.5w,0.5h}{v2}
\fmf{wiggly}{v3,v2,v1}
\fmfdot{v2}
\end{fmfgraph}
}}&\equiv& \frac{1}{2M\hbar}\int_0^{\hbar} d\otau\,p^2(\otau), \\
\label{rule6}
\parbox{30pt}{\centerline{
\begin{fmfgraph}(20,8)
\setval
\fmfleft{v1}
\fmfright{v3}
\fmfforce{0.5w,0.5h}{v2}
\fmf{plain}{v3,v2,v1}
\fmfdot{v2}
\end{fmfgraph}
}}&\equiv& \frac{1}{2\hbar}M\omega^2\int_0^{\hbar} d\otau\,x^2(\otau).
\end{eqnarray}
Feynman diagrams are built up by attaching the legs of such subgraphs to others 
or by connecting a leg with a suitable current. Note that only combinations of different 
types of subgraphs lead to non-vanishing contributions, since the connection of subgraphs
of same type,
\begin{equation}
\parbox{30pt}{\centerline{
\begin{fmfgraph}(30,8)
\setval
\fmfleft{v1}
\fmfright{v4}
\fmfforce{1/4w,0.5h}{v2}
\fmfforce{3/4w,0.5h}{v3}
\fmf{wiggly}{v4,v3,v2,v1}
\fmfdot{v2,v3}
\end{fmfgraph}
}}\,,\quad
\parbox{30pt}{\centerline{
\begin{fmfgraph}(30,8)
\setval
\fmfleft{v1}
\fmfright{v4}
\fmfforce{1/4w,0.5h}{v2}
\fmfforce{3/4w,0.5h}{v3}
\fmf{plain}{v4,v3,v2,v1}
\fmfdot{v2,v3}
\end{fmfgraph}
}}\,, 
\end{equation}
sets up a new subgraph which contains a propagator (\ref{rule0a}) or (\ref{rule0b}),
respectively. These propagators, however, are independent of $\tau$, such that 
the $\tau$-integrals related to the vertices in these subgraphs are trivial. Thus
there does not really exist a connection between these vertices and the propagators
(\ref{rule0a}) and (\ref{rule0b}) can be expressed by the currents (\ref{rule2})
and (\ref{rule3}):
\begin{eqnarray}
\label{rule7a}
\parbox{60pt}{\centerline{
\begin{fmfgraph*}(20,8)
\setval
\fmfleft{v1}
\fmfright{v2}
\fmf{wiggly}{v2,v1}
\fmflabel{$\otau_1$}{v1}
\fmflabel{$\otau_2$}{v2}
\end{fmfgraph*}
}} &=&
\parbox{70pt}{\centerline{
\begin{fmfgraph*}(30,8)
\setval
\fmfleft{v1}
\fmfforce{1/3w,1/2h}{v2}
\fmfforce{2/3w,1/2h}{v3}
\fmfright{v4}
\fmf{wiggly}{v2,v1}
\fmf{wiggly}{v4,v3}
\fmflabel{$\otau_1$}{v1}
\fmflabel{$\otau_2$}{v4}
\vertex{v2}
\vertex{v3}
\end{fmfgraph*}
}},\\
\label{rule7b}
\parbox{60pt}{\centerline{
\begin{fmfgraph*}(20,8)
\setval
\fmfleft{v1}
\fmfright{v2}
\fmf{plain}{v2,v1}
\fmflabel{$\otau_1$}{v1}
\fmflabel{$\otau_2$}{v2}
\end{fmfgraph*}
}}&=&
\parbox{70pt}{\centerline{
\begin{fmfgraph*}(30,8)
\setval
\fmfleft{v1}
\fmfforce{1/3w,1/2h}{v2}
\fmfforce{2/3w,1/2h}{v3}
\fmfright{v4}
\fmf{plain}{v2,v1}
\fmf{plain}{v4,v3}
\fmflabel{$\otau_1$}{v1}
\fmflabel{$\otau_2$}{v4}
\vertex{v2}
\vertex{v3}
\end{fmfgraph*}
}}.
\end{eqnarray}
Therefore, for $n>1$, connected diagrams containing propagators of type (\ref{rule0a}) or 
(\ref{rule0b}) {\it must} break into non-connected parts which we are not interested in. 
Analytically, this is seen by considering for example
\begin{equation}
\mean{x(\otau_1)x(\otau_2)}=\mean{\tx(\otau_1)\tx(\otau_2)}+\mean{x(\otau_1)}\mean{x(\otau_2)}.
\end{equation}
The first term on the right-hand side vanishes due to Eq.~(\ref{eq:corr1}), 
while the second simply yields $x_0^2$, which proves Eq.~(\ref{rule1b}). 
This means that only Feynman diagrams which consist of a mixture of subgraphs (\ref{rule5})
and (\ref{rule6}) contribute to the effective classical Hamiltonian. To illustrate this,
we discuss the first and second order of expansion (\ref{eq:scalham}) in more detail.

The Feynman diagrams of the first-order contribution to the effective classical Hamiltonian 
are simply constructed from the subgraphs
\begin{eqnarray}
\label{first}
&&\hspace{-20pt}H_{\omega,{\rm eff}}^{(1)}(p_0,x_0)\propto
\parbox{30pt}{\centerline{
\begin{fmfgraph}(20,8)
\setval
\fmfleft{v1}
\fmfright{v3}
\fmfforce{0.5w,0.5h}{v2}
\fmf{wiggly}{v3,v2,v1}
\fmfdot{v2}
\end{fmfgraph}
}}+
\parbox{30pt}{\centerline{
\begin{fmfgraph}(20,8)
\setval
\fmfleft{v1}
\fmfright{v3}
\fmfforce{0.5w,0.5h}{v2}
\fmf{plain}{v3,v2,v1}
\fmfdot{v2}
\end{fmfgraph}
}}\nonumber\\
&&=\frac{1}{2M\hbar}
\parbox{25pt}{\centerline{
\begin{fmfgraph}(15,15)
\setval
\fmfleft{v1}
\fmfi{wiggly}{reverse fullcircle scaled 1w shifted (0.5w,0.5h)}
\fmfdot{v1}
\end{fmfgraph}
}}
+\frac{1}{2\hbar}M\omega^2
\parbox{25pt}{\centerline{
\begin{fmfgraph}(15,15)
\setval
\fmfleft{v1}
\fmfi{plain}{reverse fullcircle scaled 1w shifted (0.5w,0.5h)}
\fmfdot{v1}
\end{fmfgraph}
}}=\frac{1}{2M\hbar}
\parbox{35pt}{\centerline{
\begin{fmfgraph}(25,8)
\setval
\fmfleft{v1}
\fmfforce{0.5w,0.5h}{v2}
\fmfright{v3}
\fmf{wiggly}{v3,v2,v1}
\fmfdot{v2}
\vertex{v1}
\vertex{v3}
\end{fmfgraph}
}}+\frac{1}{2\hbar}M\omega^2
\parbox{35pt}{\centerline{
\begin{fmfgraph}(25,8)
\setval
\fmfleft{v1}
\fmfforce{0.5w,0.5h}{v2}
\fmfright{v3}
\fmf{plain}{v3,v2,v1}
\fmfdot{v2}
\vertex{v1}
\vertex{v3}
\end{fmfgraph}
}}\nonumber\\
&&=\frac{p_0^2}{2M}+\frac{1}{2}M\omega^2x_0^2,
\end{eqnarray}
where we have used the identities (\ref{rule7a}) and (\ref{rule7b}) in the second expression
of the second line. Note that the first order term (\ref{first}) 
obviously reproduces the classical 
Hamiltonian. This is the consequence of the high-temperature expansion (\ref{eq:binham}),
since only the first-order contribution is nonzero in the limit $\beta=1/k_BT\to 0$. 
The second-order contribution reads
\begin{eqnarray}
\label{second}
H_{\omega,{\rm eff}}^{(2)}(p_0,x_0)&\propto& 
(\parbox{30pt}{\centerline{
\begin{fmfgraph}(20,8)
\setval
\fmfleft{v1}
\fmfright{v3}
\fmfforce{0.5w,0.5h}{v2}
\fmf{wiggly}{v3,v2,v1}
\fmfdot{v2}
\end{fmfgraph}
}}+
\parbox{30pt}{\centerline{
\begin{fmfgraph}(20,8)
\setval
\fmfleft{v1}
\fmfright{v3}
\fmfforce{0.5w,0.5h}{v2}
\fmf{plain}{v3,v2,v1}
\fmfdot{v2}
\end{fmfgraph}
}})
(\parbox{30pt}{\centerline{
\begin{fmfgraph}(20,8)
\setval
\fmfleft{v1}
\fmfright{v3}
\fmfforce{0.5w,0.5h}{v2}
\fmf{wiggly}{v3,v2,v1}
\fmfdot{v2}
\end{fmfgraph}
}}+
\parbox{30pt}{\centerline{
\begin{fmfgraph}(20,8)
\setval
\fmfleft{v1}
\fmfright{v3}
\fmfforce{0.5w,0.5h}{v2}
\fmf{plain}{v3,v2,v1}
\fmfdot{v2}
\end{fmfgraph}
}})
\nonumber\\
&=&-\frac{\omega^2}{8\hbar^2\beta}\left( 
8
\parbox{50pt}{\centerline{
\begin{fmfgraph}(30,8)
\setval
\fmfleft{v1}
\fmfright{v4}
\fmfforce{1/4w,0.5h}{v2}
\fmfforce{3/4w,0.5h}{v3}
\fmf{plain}{v4,v3}
\fmf{dashes_arrow}{v2,v3}
\fmf{wiggly}{v2,v1}
\fmfdot{v2,v3}
\vertex{v1}
\vertex{v4}
\end{fmfgraph}
}}
+4
\parbox{35pt}{\centerline{
\begin{fmfgraph}(15,15)
\setval
\fmfleft{v1}
\fmfright{v2}
\fmf{dashes_arrow,left=1}{v1,v2}
\fmf{dashes_arrow,right=1}{v1,v2}
\fmfdot{v1,v2}
\end{fmfgraph}
}}
\right).
\end{eqnarray}
The chain diagram is zero, while the loop diagram has the value $-\hbar^4\zeta(2)/2\pi^2$,
where 
\begin{equation}
\label{zeta}
\zeta(z)=\sum_{n=1}^\infty \frac{1}{n^z}
\end{equation} 
is the $\zeta$-function. Thus we obtain 
$H^{(2)}_{\omega,{\rm eff}}(p_0,x_0)=\beta \hbar^2\omega^2\zeta(2)/4\pi^2$.
This second-order contribution (\ref{second}) shows the characteristic types of 
Feynman diagrams in each order $n>1$
of the expansion (\ref{eq:scalham}) for the harmonic oscillator: chain and loop diagrams.
In order to calculate the $n$th-order contribution, we must evaluate these diagrams more 
general. By constructing Feynman diagrams from the product of $n$ sums of subgraphs,
\begin{equation}
H_{\omega,{\rm eff}}^{(n)}(p_0,x_0)\propto 
\underbrace{
(\parbox{30pt}{\centerline{
\begin{fmfgraph}(20,8)
\setval
\fmfleft{v1}
\fmfright{v3}
\fmfforce{0.5w,0.5h}{v2}
\fmf{wiggly}{v3,v2,v1}
\fmfdot{v2}
\end{fmfgraph}
}}+
\parbox{30pt}{\centerline{
\begin{fmfgraph}(20,8)
\setval
\fmfleft{v1}
\fmfright{v3}
\fmfforce{0.5w,0.5h}{v2}
\fmf{plain}{v3,v2,v1}
\fmfdot{v2}
\end{fmfgraph}
}})
(\parbox{30pt}{\centerline{
\begin{fmfgraph}(20,8)
\setval
\fmfleft{v1}
\fmfright{v3}
\fmfforce{0.5w,0.5h}{v2}
\fmf{wiggly}{v3,v2,v1}
\fmfdot{v2}
\end{fmfgraph}
}}+
\parbox{30pt}{\centerline{
\begin{fmfgraph}(20,8)
\setval
\fmfleft{v1}
\fmfright{v3}
\fmfforce{0.5w,0.5h}{v2}
\fmf{plain}{v3,v2,v1}
\fmfdot{v2}
\end{fmfgraph}
}})
\cdots
(\parbox{30pt}{\centerline{
\begin{fmfgraph}(20,8)
\setval
\fmfleft{v1}
\fmfright{v3}
\fmfforce{0.5w,0.5h}{v2}
\fmf{wiggly}{v3,v2,v1}
\fmfdot{v2}
\end{fmfgraph}
}}+
\parbox{30pt}{\centerline{
\begin{fmfgraph}(20,8)
\setval
\fmfleft{v1}
\fmfright{v3}
\fmfforce{0.5w,0.5h}{v2}
\fmf{plain}{v3,v2,v1}
\fmfdot{v2}
\end{fmfgraph}
}})}_{n\;{\rm times}},
\end{equation} 
it turns out that only following chain and loop diagrams contribute:\\
\begin{minipage}{5.8cm}
\begin{eqnarray}
&&\parbox{105pt}{\centerline{
\begin{fmfgraph}(100,8)
\setval
\fmfforce{0w,1/2h}{v1}
\fmfforce{10/100w,1/2h}{v2}
\fmfforce{30/100w,1/2h}{v3}
\fmfforce{50/100w,1/2h}{v4}
\fmfforce{70/100w,1/2h}{v5}
\fmfforce{90/100w,1/2h}{v6}
\fmfforce{1w,0.5h}{v7}
\fmf{wiggly}{v1,v2}
\fmf{dashes_arrow}{v3,v2}
\fmf{dashes_arrow}{v3,v4}
\fmf{dots}{v4,v5}
\fmf{dashes_arrow}{v6,v5}
\fmf{wiggly}{v6,v7}
\fmfdot{v2,v3,v4,v5,v6}
\vertex{v1}
\vertex{v7}
\end{fmfgraph}
}},\nonumber\\
&&\parbox{105pt}{\centerline{
\begin{fmfgraph}(100,8)
\setval
\fmfforce{0w,1/2h}{v1}
\fmfforce{10/100w,1/2h}{v2}
\fmfforce{30/100w,1/2h}{v3}
\fmfforce{50/100w,1/2h}{v4}
\fmfforce{70/100w,1/2h}{v5}
\fmfforce{90/100w,1/2h}{v6}
\fmfforce{1w,0.5h}{v7}
\fmf{plain}{v1,v2}
\fmf{dashes_arrow}{v3,v2}
\fmf{dashes_arrow}{v3,v4}
\fmf{dots}{v4,v5}
\fmf{dashes_arrow}{v6,v5}
\fmf{plain}{v6,v7}
\fmfdot{v2,v3,v4,v5,v6}
\vertex{v1}
\vertex{v7}
\end{fmfgraph}
}},\nonumber\\
&&\parbox{105pt}{\centerline{
\begin{fmfgraph}(100,8)
\setval
\fmfforce{0w,1/2h}{v1}
\fmfforce{10/100w,1/2h}{v2}
\fmfforce{30/100w,1/2h}{v3}
\fmfforce{50/100w,1/2h}{v4}
\fmfforce{70/100w,1/2h}{v5}
\fmfforce{90/100w,1/2h}{v6}
\fmfforce{1w,0.5h}{v7}
\fmf{plain}{v1,v2}
\fmf{dashes_arrow}{v3,v2}
\fmf{dashes_arrow}{v3,v4}
\fmf{dots}{v4,v5}
\fmf{dashes_arrow}{v6,v5}
\fmf{wiggly}{v6,v7}
\fmfdot{v2,v3,v4,v5,v6}
\vertex{v1}
\vertex{v7}
\end{fmfgraph}
}},\nonumber
\end{eqnarray}
\end{minipage}\hfill
\begin{minipage}{5.8cm}
\begin{equation}
\parbox{60pt}{\centerline{
\begin{fmfgraph}(40,40)
\setval
\fmfsurroundn{v}{8}\fmfdotn{v}{8}
\fmf{dashes_arrow,right=0.2}{v1,v2}
\fmf{dashes_arrow,left=0.2}{v1,v8}
\fmf{dashes_arrow,left=0.2}{v3,v2}
\fmf{dashes_arrow,right=0.2}{v3,v4}
\fmf{dashes_arrow,left=0.2}{v5,v4}
\fmf{dashes_arrow,right=0.2}{v5,v6}
\fmf{dashes_arrow,left=0.2}{v7,v6}
\fmf{dots,right=0.2}{v7,v8}
\end{fmfgraph}
}}.
\end{equation}
\end{minipage}\\[0.3cm]
The evaluation of the chain diagrams is easily done and yields zero. An explicit calculation
in Fourier space shows that there occur Kronecker-$\delta$'s $\delta_{m\,0}$. Since the
Matsubara sum of the Green 
function~(\ref{eq:prop})) does not contain the zero mode $m=0$, all chain diagrams are zero.

Determining the values of loop diagrams is more involved. It is obvious that loop diagrams
can only be constructed in {\it even} order ($n=2,4,6,\ldots$), since for a loop diagram
with mixed propagators (\ref{rule1a}) or (\ref{rule1b}) pairs of different subgraphs 
(\ref{rule5}) and (\ref{rule6}) are necessary. Thus we have found the result
that {\it odd} orders of expansion (\ref{eq:binham}) {\it vanish}, and only loop diagrams
for $n\in \{2,4,6,\ldots\}$ must be calculated.  
Evaluating loop diagrams of $n$th order in Fourier space is straightforward and
entails
\begin{equation}
\label{eq:loop}
\parbox{60pt}{\centerline{
\begin{fmfgraph}(40,40)
\setval
\fmfsurroundn{v}{8}\fmfdotn{v}{8}
\fmf{dashes_arrow,right=0.2}{v1,v2}
\fmf{dashes_arrow,left=0.2}{v1,v8}
\fmf{dashes_arrow,left=0.2}{v3,v2}
\fmf{dashes_arrow,right=0.2}{v3,v4}
\fmf{dashes_arrow,left=0.2}{v5,v4}
\fmf{dashes_arrow,right=0.2}{v5,v6}
\fmf{dashes_arrow,left=0.2}{v7,v6}
\fmf{dots,right=0.2}{v7,v8}
\end{fmfgraph}
}}=2\,(-1)^k\left(\frac{\hbar^2}{2\pi} \right)^{2k}\zeta(2k),
\end{equation}  
where $k=n/2$. The high-temperature expansion for the effective Hamiltonian of the harmonic
oscillator can thus be written as
\begin{equation}
\label{eq:fin1}
H_{\omega,{\rm eff}}(p_0,x_0)=\frac{p_0^2}{2M}+\frac{1}{2}M\omega^2x_0^2+\sum\limits_{k=1}^\infty\,
\beta^{2k-1}\frac{(-1)^{k+1}}{k}\left(\frac{\hbar\omega}{2\pi} \right)^{2k}\zeta(2k).
\end{equation}
Substituting the $\zeta$-function by its definition (\ref{zeta}) 
and exchanging the summations, the
last term in Eq.~(\ref{eq:fin1}) can be expressed as a logarithm
\begin{equation}
\sum\limits_{k=1}^\infty\,
\beta^{2k-1}\frac{(-1)^{k+1}}{k}\left(\frac{\hbar\omega}{2\pi} \right)^{2k}\zeta(2k)=
\frac{1}{\beta}{\rm ln}\,\left(
\prod\limits_{n=1}^\infty\left[1+\frac{\hbar^2\beta^2\omega^2}{4\pi^2n^2} \right] \right).
\end{equation}
Applying the relation
\begin{equation}
\frac{1}{z}\sinh\, z=\prod\limits_{n=1}^\infty\left(1+\frac{z^2}{n^2\pi^2} \right),
\end{equation}
we find the more familiar form of the effective classical Hamiltonian for a harmonic oscillator
\begin{equation}
\label{eq:fin2}
H_{\omega,{\rm eff}}(p_0,x_0)=\frac{p_0^2}{2M}+\frac{1}{2}M\omega^2x_0^2-\frac{1}{\beta}
{\rm ln}\,\frac{\hbar\omega\beta}{2\sinh\,\hbar\omega\beta/2}.
\end{equation}
Performing the $x_0$- and $p_0$-integrations in Eq.~(\ref{partfunc}), we obtain the 
well-known form of the partition function of the harmonic oscillator 
$Z_\omega=1/\sinh\,\hbar\omega\beta$. 
\section{Summary}
We have used a perturbative definition of the path integral in phase space 
representation which reproduces the effective classical Hamiltonian for the harmonic
oscillator.  Our procedure is an alternative way to evaluate path integrals:
The unperturbed system is trivial and the calculation of appropriate
Feynman diagrams is simple.  
Furthermore it turns out that the perturbative expansion for the effective 
classical Hamiltonian is identical to the high-temperature expansion.
\section*{Acknowledgements}
I deeply thank Professor H. Kleinert for many exciting discussions, useful hints,
and expert advice in the surrounding of my PhD studies. Also, I'm indepted
to Dr. A. Pelster and Dr. A. Chervyakov for conversations regarding the pertubatively 
defined path integral.
Finally, I'm grateful to the Studienstiftung des deutschen Volkes for support.

\end{fmffile}

\begin{thebibliography}{99}
\bibitem{PI} H. Kleinert, {\it Path Integrals in Quantum Mechanics, Statistics and 
Polymer Physics}, 2nd ed. (World Scientific, Singapore, 1995).
\bibitem{cherv} H. Kleinert, A. Chervyakov, {\it Phys. Lett. A} {\bf 269}, 63 (2000).
\bibitem{corr} H. Kleinert, A. Pelster, M. Bachmann, {\it Phys. Rev. E} {\bf 60}, 
2510 (1999).
\bibitem{mag} M. Bachmann, H. Kleinert, A.Pelster, {\it Phys. Rev. A} {\bf 62}, 52509 (2000);
M. Bachmann, H. Kleinert, A. Pelster, {\it Phys. Lett. A}  {\bf 279}, 23 (2001).
\end{thebibliography}
\end{document}